\documentclass{aa}  

\usepackage[varg]{txfonts}
\usepackage{graphicx}
\usepackage{graphics}
\usepackage{rotating}
\usepackage{pdflscape}
\usepackage{lscape}
\usepackage{longtable}
\usepackage{enumerate}
\usepackage[flushleft]{threeparttable}
\usepackage{ulem}

\usepackage{natbib,twoopt}
\bibpunct{(}{)}{;}{a}{}{,}             
\begin{document}

   \title{The Gaia-ESO Survey: the first abundance determination of the pre-main-sequence cluster Gamma Velorum\thanks{Based on observations collected at the ESO telescopes under programme 188.B3002, the Gaia-ESO large public spectroscopic survey.},\thanks{Full Tables 1, 2, 3 and 4 are only available in electronic form at the CDS via anonymous ftp to cdsarc.u-strasbg.fr (130.79.128.5) or via http://cdsweb.u-strasbg.fr/cgi-bin/qcat?J/A+A/.}}

\author{L. Spina\inst{1,2}, S. Randich\inst{1}, F. Palla\inst{1}, G.~G. Sacco\inst{1}, L. Magrini\inst{1}, E. Franciosini\inst{1}, L. Morbidelli\inst{1}, L. Prisinzano\inst{3}, E. J. Alfaro\inst{4}, K. Biazzo\inst{5}, A. Frasca\inst{5}, J.~I. Gonz\'{a}lez Hern\'{a}ndez\inst{6,7}, S.~G. Sousa\inst{8,9}, V. Adibekyan\inst{8}, E. Delgado-Mena\inst{8}, D. Montes\inst{10}, H. Tabernero\inst{10}, A. Klutsch\inst{5}, G. Gilmore\inst{11}, S. Feltzing\inst{12}, R.~D. Jeffries\inst{13}, G. Micela\inst{3}, A. Vallenari\inst{14}, T. Bensby\inst{11}, A. Bragaglia\inst{15}, E. Flaccomio\inst{13}, S. Koposov\inst{10},  A.~C. Lanzafame\inst{16}, E. Pancino\inst{15}, A. Recio-Blanco\inst{18}, R. Smiljanic\inst{19},\inst{20}, M.~T. Costado\inst{4}, F. Damiani\inst{3}, V. Hill\inst{18}, A. Hourihane\inst{10}, P. Jofr\'{e}\inst{10}, P. de Laverny\inst{18}, T. Masseron\inst{10}, C. Worley\inst{11}}
\offprints{L. Spina}

\institute{
INAF--Osservatorio Astrofisico di Arcetri, Largo E. Fermi, 5, I-50125 Firenze, Italy \email{lspina@arcetri.astro.it}
\and
Universit\`{a} degli Studi di Firenze, Dipartimento di Fisica e Astrofisica, Sezione di Astronomia, Largo E. Fermi, 2, I-50125, Firenze, Italy
\and
INAF - Osservatorio Astronomico di Palermo, Piazza del Parlamento 1, 90134, Palermo, Italy
\and
Instituto de Astrof\'{i}sica de Andaluc\'{i}a-CSIC, Apdo. 3004, 18080 Granada, Spain
\and
INAF--Osservatorio Astrofisico di Catania, via S. Sofia, 78, I-95123 Catania, Italy
\and
Instituto de Astrofisica de Canarias (IAC), E-38205 La Laguna, Tenerife, Spain
\and
Depto. Astrofisica, Universidad de La Laguna (ULL), E-38206 La Laguna, Tenerife, Spain
\and
Centro de Astrofisica, Universidade do Porto, Rua das Estrelas, 4150-762, Porto, Portugal
\and
Departamento de F\'isica e Astronomia, Faculdade de Ci\^encias, Universidade do Porto, Rua do Campo Alegre, 4169-007 Porto, Portugal
\and
Departamento de Astrofisica, Universidad Complutense de Madrid (UCM), Spain
\and
Institute of Astronomy, University of Cambridge, Madingley Road, Cambridge CB3 0HA, United Kingdom
\and
Lund Observatory, Department of Astronomy and Theoretical Physics, Box 43, SE-221 00 Lund, Sweden
\and
Astrophysics Group, Research Institute for the Environment, Physical Sciences and Applied Mathematics, Keele University, Keele, Staffordshire ST5 5BG, United Kingdom
\and
INAF - Padova Observatory, Vicolo dell'Osservatorio 5, 35122 Padova, Italy
\and
INAF - Osservatorio Astronomico di Bologna, via Ranzani 1, 40127, Bologna, Italy
\and
Dipartimento di Fisica e Astronomia, Sezione Astrofisica, Universit\'{a} di Catania, via S. Sofia 78, 95123, Catania, Italy
\and
ASI Science Data Center, Via del Politecnico SNC, 00133 Roma, Italy
\and
Universit\'{e} de Nice Sophia Antipolis, CNRS, Observatoire de la C\^{o}te d'Azur,  BP 4229, F-06304, Nice Cedex 4, France
\and
Department for Astrophysics, Nicolaus Copernicus Astronomical Center, ul. Rabia\'{n}ska 8, 87-100 Toru\'{n}, Poland
\and
European Southern Observatory, Karl-Schwarzschild-Str. 2, 85748 Garching bei M\"{u}nchen, Germany
}

 \date{Received 29 January 2014; Accepted 25 April 2014}

  \abstract
{Knowledge of the abundance distribution of star forming regions and young 
clusters is critical to investigate a variety of issues, 
from triggered star formation and chemical enrichment by nearby supernova explosions to the ability to form planetary systems.
In spite of this, detailed abundance studies are currently available
for relatively few regions.}
{In this context, we present the analysis of the metallicity 
of the Gamma Velorum cluster, based on the products distributed
in the first internal release of the Gaia-ESO Survey.
}
{The Gamma Velorum candidate members have been observed with FLAMES,
using both UVES and Giraffe, depending on the target
brightness and spectral type. In order to derive a solid metallicity 
determination for the cluster, membership of the observed stars must 
be first assessed.
To this aim, we use several 
membership criteria including radial velocities, surface gravity estimates, and the detection of the photospheric lithium line.}
{
Out of the 80 targets observed with UVES, we identify 14 high-probability
members. 
We find that the metallicity of the cluster is slightly subsolar, with 
a mean [Fe/H]=$-$0.057$\pm$0.018~dex. Although J08095427-4721419 is one of the high-probability members, its metallicity is significantly larger than the cluster average. We speculate about its origin as the result of recent accretion episodes of rocky bodies of $\sim$60~$M_{\oplus}$ 
hydrogen-depleted material from the circumstellar disk.
}
{}
\authorrunning{L. Spina et al.}
\titlerunning{Gaia-ESO Survey: the first abundance determination of the Gamma Velorum cluster}
\maketitle
\section{Introduction}
Open clusters are excellent tracers of the chemical pattern of the Galactic thin disk and its evolution (e.g., \citealt{Friel95}). The youngest clusters, the so-called pre-main-sequence (PMS) clusters with ages $\lesssim$~50~Myr, are of particular interest since they are still close to their birthplaces and contain a homogeneous stellar population that has not had time to disperse through the Galactic disk. Thus, they are key objects to trace the current chemical composition of the solar neighborhood and its evolution in space and time. 

Furthermore, and more specifically,
determination of the chemical content of young clusters and star forming regions (SFRs) is critical for a variety of reasons that we summarize below. 
First, as originally discussed in the series of papers by Cunha and 
collaborators, knowledge of the abundance pattern
allows us to investigate the common origin of different subgroups in a given association and it sheds light on the possible presence of enrichment caused by the explosion of a nearby supernova (\citealt[and references therein]{Cunha98,Biazzo11a}). Indeed, in the triggered star formation scenario, 
newly formed massive stars belonging to a first generation of stars in a giant
molecular cloud and ending their lifetime with supernova (SN) explosions,
disperse the parent molecular cloud, preventing further star formation to occur in the immediate surroundings. At the same time, however, winds and SN-driven
shock waves are
thought to trigger new star formation events at larger distances; 
since supernovae are major nucleosynthesis sites, these explosions, 
may also chemically enrich parts of the surrounding interstellar gas, and hence
the newly formed second generation of stars
(\citealt{Cunha92,Cunha94} and references therein). 
Finding direct evidence of such selective enrichment in young clusters and SFRs would clearly give insights into a process that has occurred innumerable times in the past, not just in our own galaxy.

In addition, as in the case of old populations, the metal content of PMS clusters is a critical parameter for the determination of their distance, age, and individual stellar masses of their members. Metallicity has an effect on the internal stellar structure and on the surface properties through opacity: 
even relatively minor changes in the metal content could imply that there are differences in the derived cluster ages,
distances, and masses \citep{sherry08}. These parameters in turn are critical for the determination of the initial mass function (IMF) and the star formation history within each region, as well as for investigating 
different properties such as disk lifetimes and the rotational evolution of young stars.

Third, recent theoretical studies have suggested that metallicity has an important impact on the evolution of circumstellar disks and their ability to form planets. For example, \citet{ercolano10} have shown that disks should dissipate quickly in a metal-poor environment. Observational studies on the disk lifetime at low-metallicity are controversial. On the one hand,
support for the theoretical predictions
has been provided by \citet{Yasui10} who found that the disk fraction (f$_d$) in low-metallicity clusters (with [O/H]$\sim$$-0.7$) declines rapidly and approaches f$_d\sim$10$\%$ in $\lesssim$1~Myr, significantly earlier than solar-metallicity clusters for which the timescale is $\sim$5-7~Myr \citep{SiciliaAguilar06,Mordasini12}.
On the other hand, based on Hubble Space Telescope mass accretion rate measurements, \citet{Spezzi12} suggest that disks in metal-poor clusters of the Large Magellanic Cloud may be long lived with respect to the Milky Way.

Finally, it is worth mentioning the 
correlation between metallicity
and elemental abundances and the presence of giant 
planets around old
solar-type stars \citep{Gonzalez98,Santos04,Johnson10}. 
In particular, \citet{Gilli06}, \citet{Neves09}, \citet{Kang11} and \citet{Adibekyan12a,Adibekyan12b} have shown that the chemical differences between stars with and without exoplanets are not limited to the iron content, but also to the abundance of some refractory elements (e.g., Mg, Al, Sc, Ti, V and Co). Thus, studying the metal content of
nearby young clusters, hosting a number of T-Tauri stars 
with circumstellar disks and likely on the verge of forming planets, 
may provide useful constraints
to studies of planet formation scenarios and their 
timescales.

In spite of all these exciting aspects, 
relatively few studies have addressed the issue of the metal content of 
PMS clusters and SFRs (see, e.g., \citealt{James06,Santos08,
DOrazi09,Biazzo11a,BIazzo11b}), rather mostly focusing on well 
studied, nearby regions like Orion and Taurus-Auriga.
A metallicity close to or slightly lower than the solar value has been measured for all these
regions; interestingly, and at variance with older clusters,
none of them appears to be metal-rich \citep{Biazzo11a}. Since, as mentioned, only relatively few young clusters
and only very few stars per region have high resolution abundance measurements,
additional studies are clearly warranted.

The Gaia-ESO Survey \citep{Gilmore12,Randich13} is a large public spectroscopic survey observing all the components of the Galaxy (bulge, thin and thick disks, and halo). The project makes use of the FLAMES spectrograph mounted at the VLT to obtain spectra of about 10$^5$ stars, including candidate members of 90-100
open clusters. This large sample of observations will allow us to accurately study of the kinematical and chemical abundance distributions in the Milky Way and also to fully sample the age-metallicity-mass/density-Galactocentric distance parameter space within the open clusters selected. In this framework, the Gaia-ESO Survey represents a unique opportunity not only to extend the sample of young clusters and star forming regions with metallicity and abundance determinations, 
but also to perform a homogeneous study based on a large stellar sample within each region.
The Gaia-ESO Survey will provide a comprehensive and homogeneous view on the chemical contents of the youngest clusters in the Galaxy, based on the analysis of a large sample of clusters only near its completion. In these initial stages of the survey, however, studies the abundance pattern of individual clusters are very valuable, not only to test methods and tools, but also because, as mentioned, few young clusters so far have solid abundance determination. Adding information and statistics it
is hence very important.

In particular, in this paper we present the products released internally to the Gaia-ESO Survey consortium on the first observed PMS cluster:
Gamma Velorum. The cluster properties, the target selection and spectral analysis are detailed in Sect.~\ref{pipeline}. The comparison between the main stellar parameters derived with the two different spectrographs, UVES and Giraffe, is given in Sect.~$\ref{verification}$. The identification of the cluster members is presented in Sect.~$\ref{mem-analysis}$, while the results of the elemental abundance determination are 
discussed in Sect.~$\ref{Discussion}$. 
Finally, the conclusions are
outlined in Sect.~$\ref{conclusions}$.

\section{Observations and data processing\label{pipeline}}
The work presented in this paper is based on the results of the analysis of the spectra obtained during the first six months of observations (January - June 2012) and released internally in the GESviDR1Final catalog (August 2013). 
In the following, we describe the properties of Gamma Velorum, the target selection, the observations, and the spectroscopic analysis.

\subsection{The Gamma Velorum open cluster \label{opencluster}}
Gamma Velorum is a nearby ($\sim$350~pc) open cluster for
which Jeffries et al (2009) originally claimed an age of $\sim$5-10~Myr,
but that could instead be older than 10 Myr (but younger than 20 Myr;
see discussion in Jeffries et al. 2014). 
Its low-mass members are distributed around a double-lined spectroscopic binary system (hereafter $\gamma^2$~Vel, as in \citealt{Jeffries09}), 
composed of a Wolf-Rayet (hereafter WR) WC8 star (the closest Wolf-Rayet star to the Sun; \citealt{Smith68}) and an O8 massive star \citep{Schaerer97}. 
\citet{Pozzo00} first recognized the presence of 
low-mass stars around the more massive objects.
Because of the low extinction and reddening (A$_V$=0.131 and $E_{B-V}$=0.038 ; Jeffries et al. 2009), the sparse disk population and youth of the Gamma Velorum association, the sequence  of the cluster is clearly visible in the optical color-magnitude diagrams presented by 
Jeffries et al. (2009).

On a larger scale Gamma Velorum lies in the so-called Vela complex (see \citealt{Pettersson08,Sushch11}), a very composite region characterized, {\it inter alia}, by the presence of a number of PMS clusters (e.g, Gamma Velorum, Tr 10, and NGC 2547), three OB associations \citep{Humphreys78,Brandt71,Slawson88} and two supernova remnants (the Gum Nebula and the Vela SNR). The latter have been created by two or more supernovae explosions that occurred 1-6~Myr and 11400~yr ago \citep{Pettersson08}. The shocks from the latter SN have not yet reached the Gamma Velorum cluster \citep{Sushch11}, but it is clear that the environment has been subject to a fast dynamical evolution. In this context, the analysis of the Gaia-ESO Survey data has led \citet{Jeffries14} to conclude that 208 members of the Gamma Velorum cluster, targeted by Giraffe and identified through their lithium content, are grouped in two distinct kinematic populations. More specifically,
through a maximum-likelihood fit of the RV distribution they
have found that the first kinematic component (population A), 
centered at R$V_{1}$$=$16.70~km/s, is narrower and consistent with virial equilibrium ($\sigma_{1}$$=$0.28~km/s), while the second component (population B)
is much broader ($\sigma_{2}$$=$1.85~km/s) and centered at higher 
velocities, i.e., R$V_{2}$$=$18.58~km/s.

Interestingly, $\gamma^2$~Vel appears to be younger than the low-mass stars. 
Indeed, the relation and interactions of $\gamma^2$ Vel 
with the low-mass cluster members is still debated, mainly because of the age of the central WR star. In fact, even if the most recent $\gamma^2$ Vel distance determinations 36$8_{-13}^{+38}$~pc \citep{Millour07}, 33$6_{-7}^{+8}$~pc \citep{North07} and 33$4_{-40}^{+32}$~pc \citep{VanLeeuwen07} support its association with the cluster, the age estimates of 3.5~$\pm$~0.4~Myr \citep{North07}, and 5.5~$\pm$~1~Myr \citep{Eldridge09} indicate that $\gamma^2$ Vel  is younger than the majority of the low-mass members of the cluster.

In spite of the remarkable properties of the cluster and of the Vela complex which makes Gamma Velorum a suitable target for a spectroscopic survey, its iron abundance is still unknown. 
Gaia-ESO Survey observations hence allow us to perform the first abundance study of this
cluster.
\subsection{Target selection and Observations}
The Gaia-ESO Survey observations are performed with the multi object optical spectrograph FLAMES on the VLT \citep{Pasquini02}. This instrument makes use
of two spectrographs, Giraffe (132 fibers) and UVES (eight fibers).
 
We based the target selection criteria on homogeneous photometric data, 
covering a large area of the cluster field. In particular, we chose the list of 
targets considering only the sources within a region of 0.9 square degrees centered on $\gamma^2$ Vel and studied by \citet{Jeffries09}. 
We selected these targets mostly following the guidelines for cluster observations (see \citealt{Bragaglia14}). 

\begin{figure}
\hspace{-0.8cm}
\includegraphics[width=0.55\textwidth]{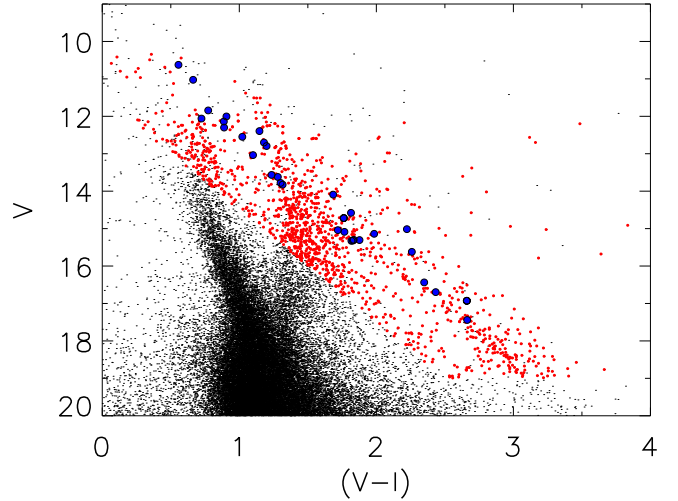}
\caption{Color-magnitude diagram of the 1283 stars observed in Gamma Velorum (in red), along with all the stars (in black) located in the field of view centered on $\gamma^2$~Vel with photometry reported in \citet{Jeffries09}. The known members from this paper are shown in blue.}
\label{cmd_all}
\end{figure}

The final sample was chosen so as to include all photometric
candidate members in a region of the color-magnitude
diagram around the cluster sequence, defined by previously known members. We considered all stars falling within $\pm$1.5~mag of the cluster sequence as
high priority targets. A few lower priority stars have also been targeted 
to use spare fibers.
The color-magnitude diagram (CMD) of the selected sources is shown in Fig.\ref{cmd_all}. The 
cluster sequence identified by \citet{Jeffries09} is clearly visible as the upper concentration of red dots with (V$-$I$)\sim$1.8$-$3.2. 
The spectroscopic survey is limited to V$\lesssim$13.5~mag  
and V$\lesssim$19~mag for UVES and Giraffe. 
Further discussion of target selection can also be found in Jeffries et al. (2014).

A total of 18 fields, forming a mosaic around $\gamma^2$~Vel, were completed during runs A (nights from 2011-12-31 to 2012-01-02) and B (night 2012-02-12) of P88, using the CD$\#$3 cross-disperser ($\lambda=$4770-6820 $\AA$; R$=$47000) for UVES and the HR15N grating ($\lambda=$6440-6820 $\AA$; R$\sim$17000) for Giraffe. Each field was observed for either 20~min (nine fields) or 50~min (nine fields). 
The exposure times hence vary from 20 min for brighter stars (V$\leq$12 and V$\leq$16 for UVES and Giraffe, respectively) to 50 min for fainter stars (12$<$V$<$13.5 and 16$<$V$<$19). Stars lying in overlapping fields have longer exposure times. We acquired spectra for a total of 1242 and 80 individual stars with Giraffe and UVES, respectively. We observed 39 stars with both spectrographs.
Signal-to-noise ratios (SNRs) for the UVES spectra are in the range 20-300, with a median value of 116. The Giraffe spectra have SNRs ranging between 3 and 300 with a median of 84.

\subsection{Available data}
The Gaia-ESO Survey is structured in 20 working groups (WGs) dedicated to different tasks.
Data reduction and determination of radial velocities (RVs) and projected rotational velocities
are carried out, independently for Giraffe and UVES, by two different teams of WG7. 
The Giraffe data are reduced using a pipeline, specifically developed for the Gaia-ESO survey,
which performs the basic steps of the data reduction process (i.e., bias subtraction, flat-fielding, spectra extraction, and
wavelength calibration), sky subtraction, and calculation of preliminary RVs and projected rotational velocities, 
by cross-correlating the spectra with a grid of templates. To improve the precision of the RVs and 
projected  rotational velocities, we fitted the reduced spectra with a low-order polynomial multiplied by a template spectrum.
The RV, the projected rotational velocities, the polynomial coefficients and the template parameters (temperature, gravity and metallicity)
are free parameters of the fit, with initial guesses derived by the first pipeline. 
We reduced the UVES data using the FLAMES-UVES ESO public pipeline.
A specific pipeline developed for the Gaia-ESO Survey is used for the sky subtraction and
the calculation of RVs and projected rotational velocities, by cross-correlating
each spectra for a grid of templates. A more detailed discussion of the procedures
used for the data reduction, and the calculation of RV, and projected rotational velocities is reported
in Lewis et al. (2014, in preparation) and \citet{Jeffries14} for Giraffe, and in \citet{Sacco14} for UVES.

As for spectrum analysis, WG11 (including the contribution of up to 13 nodes) 
is dedicated to the analysis of the UVES spectra of F-G-K stars, while WG12  (composed by four nodes) focuses on young stars, analyzing both 
UVES and Giraffe spectra.
The analysis performed by WG11 and WG12 is described in detail 
in \citet{Smiljanic14} and Lanzafame et al. (in prep), respectively. 
Whereas we will briefly describe here how the recommended parameters
released to the consortium are derived, we refer to the above two papers
for a full description of the approach and methodologies.

Both WG11 and WG12 benefit from the contribution of nodes that use different 
methods of analysis. These different approaches can be summarized as follows: i) nodes that employ the equivalent width (EW) analysis; the atmospheric parameter determination is based on the excitation and ionization balance of the iron lines; ii) nodes that use spectrum synthesis and estimated atmospheric parameters from a $\chi^{2}$ fit to observed spectra; in some cases the grid of templates is composed by observed spectra of slow-rotating, low-activity stars; iii) multi linear regression methods that simultaneously determine the stellar parameters of an observed spectrum by the projection of the spectrum onto vector functions, constructed as an optimal linear combination of the local synthetic spectra. 
The parameters released in GESviDR1 are obtained by each of the two working
groups by computing the median value of the results provided by the nodes, after the outliers have been discarded. 
Uncertainties are the node-to-node dispersions. We mention that the consortium uniformly makes use of MARCS models of stellar atmospheres \citep{Gustafsson08} that assume the solar abundances from \citet{Grevesse07}. Also, common atomic data have been used for the analysis of all the spectra of the 
Gaia-ESO Survey. 
Similarly, more than one node measure the
strength of the Li~{\sc i} line at 6707.8~$\AA$ in both Giraffe 
and UVES spectra. 
The nodes use independent methods to derive the EW of this features: 
specifically, some of them apply a Gaussian fitting to the line, while others
are based on the direct profile integration of the line. 
The median value of the EW (or the average, when only two nodes provided 
the measurement) are then adoped.
All these procedures are detailed in Lanzafame et al. (in prep).

Released parameters for Gamma Velorum include 
radial and rotational velocities, CCFs 
and the products of the spectrum analysis. The latter include 
the main atmospheric parameters ($\rm T_{\rm eff}$, $\log$~g, and [Fe/H]) and other parameters (e.g., veiling, strength of the Li~{\sc i} 
line at 6707.8~$\AA$, H$\alpha$, etc.), along with their uncertainties.
All our UVES targets, along with their with RVs and parameters, when available, are listed in Tables \ref{67parameters} and \ref{UVES_targets}.
Individual elemental abundances are also provided for UVES spectra, whenever they can be measured. The first four rows of Table~\ref{uves-summary} represent a brief outline of the data obtained from the UVES spectra. 

\begin{table*}
\centering                                      
\caption{Stellar parameters of the 80 UVES targets.}              
\label{67parameters}      
\begin{tabular}{lllllll}          
\hline\hline                        
ID & Cname & R.A. & DEC. & $\rm T_{\rm eff}$ & $\log g$ & [Fe/H] \\
 & & (J2000) & (J2000) & (K) & (dex) & (dex) \\
\hline 
1 & 08063616$-$4748206 & 08 06 36.16 & $-$47 48 20.6 & 6726$\pm$347 & 4.16$\pm$0.21 & $-$1.51$\pm$0.20 \\
2 & 08064772$-$4659492 & 08 06 47.72 & $-$46 59 49.2 & 5776$\pm$49  & 4.20$\pm$0.08 & $-$0.02$\pm$0.03 \\
... & ... & ... & ... & ... & ... & ... \\
\hline                                             
\end{tabular}
\end{table*}

\begin{table*}
\centering                                      
\caption{Quantities used for the membership analysis of the UVES sample. The table shows that we have identified eight high-probability members and eight HCMs (see text). RV values are not corrected for the 1.1~km/s systematic shift.}              
\label{UVES_targets}      
\begin{tabular}{llllllllll}          
\hline\hline                        
ID & RV & EW(Li) & (B-V$)_{0}$ & $V_{0}$ & RV & $\log$ g & Li & CMD & Final \\
 & (km/s) & (m$\AA$) & (mag) & (mag) & mem. & mem. & mem. & mem. & mem. \\
\hline 
1 & 13.5 & $<$20 & 0.13 & 1.97 & N & Y & HCM & N & N \\
2 & 10.4 & $<$10 & 0.64 & 4.99 & N & Y & N & ... & N \\
... & ... & ... & ... & ... & ... & ... & ...  & ...  & ...  \\
\hline                                             
\end{tabular}
\end{table*}

To summarize: 
\begin{itemize}
\item  80 UVES and 1242 Giraffe targets observed in the Gamma Velorum fields;
\item we found six UVES targets to be double-lined binaries (SB2; see Section 4). In the Gaia-ESO catalog, RV values are available for all of these systems. Also, the main parameters of one SB2 have been delivered by the consortium.
\item RV estimates are available for 73 of the UVES targets;
hereafter we disregard the RV values of five UVES targets either
with poor quality spectra or that are early-type stars
or fast rotating sources 
(v$\sin i$$>$100~km/s), and hence the inferred radial velocities
are highly uncertain. Moreover the RV values of the six SB2 observed with 
UVES are not considered. Thus, the final sample of stars with
available and reliable RV estimates consists of 62 UVES stars. 
The RV values for the UVES sample are listed in Table~\ref{UVES_targets}. 
We refer to \citet{Jeffries14} for the RV estimates of the Giraffe targets;
\item A measurement or an upper-limit of the EW of the Li~{\sc i} line 
is available 
for all stars with the exception of four warm stars in the UVES sample 
that do not show any Li feature in their spectra. For these four stars, 
we assume a 3$\sigma$ detection upper-limit using the Cayrel formula 
\citep{Cayrel88}. These values are listed in Table~\ref{UVES_targets}. As for the RV values, we also refer to \citet{Jeffries14} for the Li equivalent width measurements in the Giraffe spectra;
\item After the rejection of the main parameters derived for the SB2 target, 
67 stars observed with UVES have an estimate of
the fundamental parameters. Note that these are available
for 36 of 39 stars observed with both spectrographs. The main parameters 
for all the UVES targets are listed in Table~\ref{67parameters}, those obtained
from Giraffe spectra for stars observed with both spectrographs are listed in Table~\ref{uvestoo}, while the main parameters of the Giraffe members identified by \citet{Jeffries14} are listed in Table~\ref{gir_mem}.
The mean uncertainties of the parameters derived from UVES spectra are: $<$$\sigma_{Teff}$$>$$=$120~K, $<$$\sigma_{\log~\rm g}$$>$$=$0.17~dex, $<$$\sigma_{[Fe/H]}$$>$$=$0.10~dex; 
\item  Finally, individual elemental abundances have been derived for 47
stars observed with UVES.
\end{itemize}
Note that a few stars with atmospheric parameters do not have an RV estimate and {\it viceversa}.

\begin{table*}
\centering                                      
\caption{Stellar parameters of the 39 stars targeted by both UVES and Giraffe.}              
\label{uvestoo}      
\begin{tabular}{llll}          
\hline\hline                        
Star & $T_{eff}$ & log g & [Fe/H] \\
 & (K) & (dex) & (dex) \\
\hline 
08064772$-$4659492 & 5765$\pm$70 & 4.09$\pm$0.27 & $-$0.19$\pm$0.11 \\
08065592$-$4704528 & 4436$\pm$57 & 2.48$\pm$0.28 & $-$0.11$\pm$0.02 \\
... & ... & ... & ... \\
\hline                                             
\end{tabular}
\end{table*}

\begin{table*}
\centering                                      
\caption{Iron abundances of the 208 members targeted by Giraffe.}              
\label{gir_mem}      
\begin{tabular}{llllll}          
\hline\hline                        
Star & R.A. & DEC. & $T_{eff}$ & log g & [Fe/H] \\
 & (J2000) & (J2000) & (K) & (dex) & (dex) \\
\hline 
08064077$-$4736441 & 08 06 40.77 & $-$47 36 44.1 & 4798$\pm$98 & 2.77$\pm$0.25 & $-$0.18$\pm$0.02 \\
08064390$-$4731532 & 08 06 43.90 & $-$47 31 53.2 & 3259$\pm$60 & 4.76$\pm$0.15 & $-$0.32$\pm$0.17 \\
... & ... & ... & ... & ... & ... \\
\hline                                             
\end{tabular}
\end{table*}

\section{UVES vs. Giraffe\label{verification}}

As mentioned, atmospheric parameters and [Fe/H] values have been also released 
for the Giraffe targets, however, since the analysis of high-resolution spectra should
yield more reliable iron abundance values (see Sect.~5.1), 
most of our scientific analysis will focus on the results of the 
UVES observations. On the other hand,
we will mostly use the Giraffe sample as a control sample
to infer the membership of the UVES targets; therefore,
in this section we take 
advantage of the stars observed with both spectrographs to check for the 
consistency of the inferred parameters.
In particular, we will make a detailed comparison of the
RVs, lithium 
EWs, and atmospheric stellar parameters 
(T$_{\rm eff}$, $\log$~g, and [Fe/H]). 

\subsection{Radial velocities \label{verification-RV}}

\begin{figure}
\hspace{-0.8cm}
\includegraphics[width=0.55\textwidth]{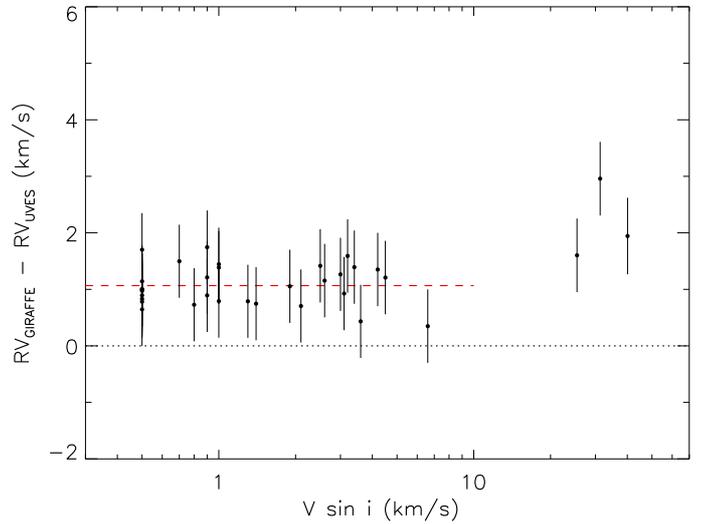}
\caption{Comparison of the RV of the 34 stars with both Giraffe and UVES spectra and available RV values. The difference of the RVs is plotted as a function of the stellar rotational velocity ($v\sin~i$). 
The red dashed line represents the offset between R$V_{\rm Giraffe}$ and R$V_{\rm UVES}$ for the 31 stars with v$\sin~i<10$~km/s.}
\label{RV_UVES_GIRAFFE}
\end{figure}

In Fig.~\ref{RV_UVES_GIRAFFE}, we show the difference between the values of the
RV as a function of the projected rotational velocity derived from the UVES spectra.
In the case of UVES, we adopt as final RV the mean of the two values obtained using the upper and lower spectral regions. 
As for the error bars, we assume the largest 
value between the error quoted in the survey catalog ($\pm$0.6~km/s) and 
the difference 
between the RVs measured independently in the two CCDs \citep{Sacco14}. 
We note that R$V_{\rm
Giraffe}$ is systematically higher than R$V_{\rm UVES}$ by 1.1$\pm$0.4~km/s 
(red dashed line in Fig.~2) up to about v$\sin i=$10~km/s and the difference
increases
for larger rotational velocities. 
While the origin of this offset needs 
further investigation (see Sacco et al. 2014), for the time being we applied an offset of $+1.1$~km/s to UVES RVs.

\subsection{Li equivalent widths}
\begin{figure}
\hspace{-0.8cm}
\includegraphics[width=0.55\textwidth]{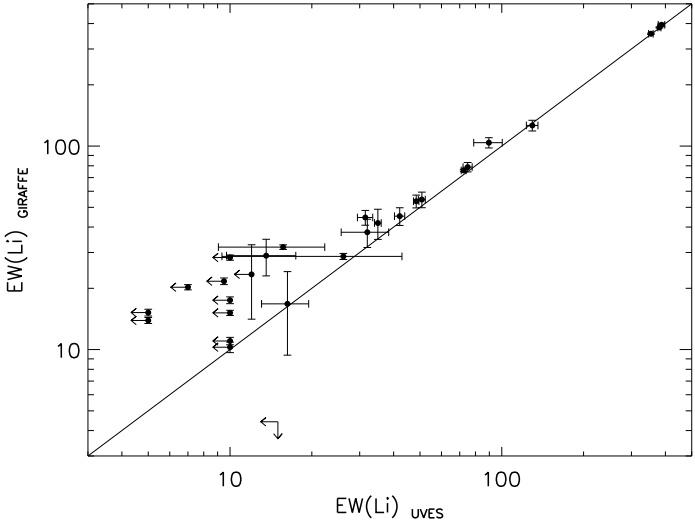}
\caption{Comparison of the 17 EW(Li) measurements obtained from both the Giraffe and UVES spectra.}
\label{GIR_UVES_EWLi}
\end{figure}

In Fig.~\ref{GIR_UVES_EWLi} we show a comparison of the
EW of the Li~{\sc i}~6707.8~\AA~line measured in Giraffe and 
UVES spectra, respectively. The figure indicates a very good agreement for
most of the stars down to about 30 m\AA; 
a discrepancy between the values is instead present below that value, 
where the Giraffe measurements are systematically higher than the UVES measurements. 
This difference needs to be further investigated and may be related to
the different resolving powers
and the blending with the nearby Fe~{\sc i}~6707.4~$\AA$ line; however, we stress
that it will not affect our 
discussion and conclusions on lithium membership, since the threshold
between Li members and nonmembers is set at higher values of the EWs
(see Sect.~4.3).

\subsection{Atmospheric parameters}
\begin{figure}
\hspace{-0.5cm}
\includegraphics[width=0.53\textwidth]{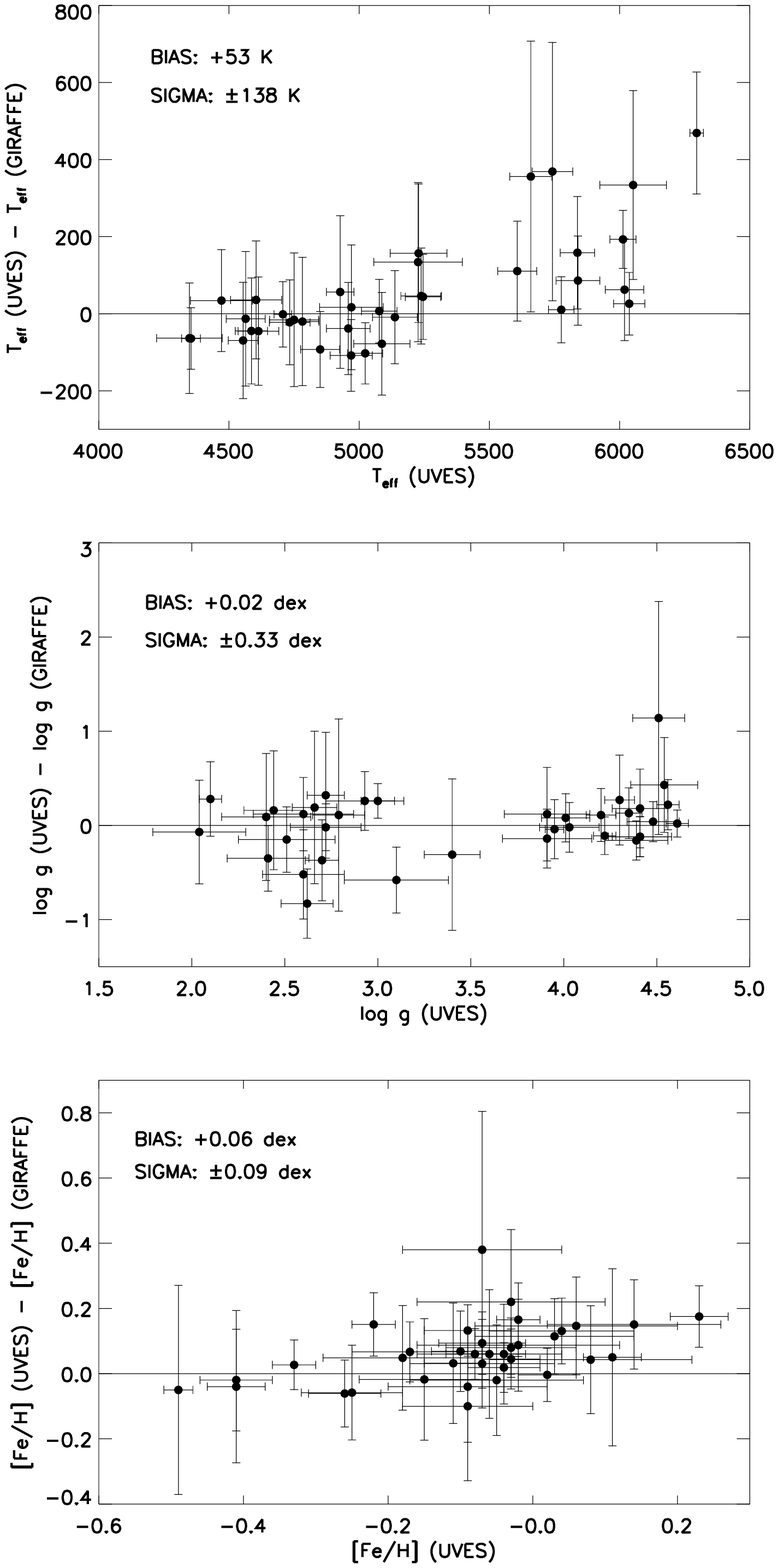}
\caption{Comparison of the stellar parameters of the stars observed with both Giraffe and UVES. From top to bottom, we show the effective temperature, the surface gravity, and the iron abundance. Systematic biases and standard deviations are reported in each panel.}
\label{GIR_UVES_mainpar}
\end{figure}

In Fig. \ref{GIR_UVES_mainpar}, we compare the fundamental parameters derived from Giraffe and UVES analyses for 36 of the 39 stars observed
with both instruments. As in the case of EW(Li), we conclude that the two spectrographs yield compatible
values within the errors for the majority of the stars. 
The only discrepancy is seen for the effective temperature 
of warm stars ($\rm T_{\rm eff}>$5500~K) for which
the Giraffe analysis gives somewhat lower values than UVES, but is still 
marginally consistent with them.
Again, the origin of these differences is under
investigation, but it does not affect our conclusions on UVES membership.
Also note that, because of the lower resolution and the shorter spectral range, 
the uncertainties on the data 
derived from Giraffe are larger. 
This widens the scatter of the data without a significant implication for 
our analysis.

To summarize, whereas we will account for the offset between Giraffe and UVES in the following
RV membership analysis, no systematic biases are present for lithium and 
$\log$~g values, the additional
two criteria that we will use for confirming the membership of UVES candidates.

\section{Membership analysis \label{mem-analysis}}
In this section, we will use the spectroscopic information, specifically,
RVs, the strength of the Li line, and the stellar surface gravity,
along 
with the position of the targets in the CMD, 
to perform the membership analysis of the UVES targets.
In Table~\ref{uves-summary}, we summarize each step of the selection procedure that, starting from the 80 stars observed with UVES, leads to a restricted sample of high-probability members that will be used for the metallicity analysis, which is the main goal of the present paper.

\begin{table*}
\caption{Summary of the selection procedure of UVES candidate members}
\begin{threeparttable}
\begin{center}
\begin{tabular}{l|ll}
\hline\hline
Observed & 80 candidates & 39 in common with Giraffe \\ \hline
Binaries   & 6 SB2 & discarded from the sample\\ \hline
RV estimates &  62 candidates & 7 RV members\\
& & 55 RV nonmembers (or SB1) \\ \hline
Fundamental parameters &  67 candidates & 37 MS/PMS stars \\
& & 30 giants \\ \hline
EW(Li) & 74 candidates $-$ 30 (giants) $=$ 44 & 8 Li members \\
excluding giants &  & 19 HCM* (8 rejected as RV nonmembers)\\
& & 17 Li nonmembers \\ \hline
CMD & 8 (Li members) + 11 (HCM) $=$ 19 & 7 Li mem. consistent with ZAMS \\
& & 1 Li mem. below ZAMS \\
& & 5 HCM consistent with ZAMS \\
& & 6 HCM below ZAMS (rejected as nonmembers)\\ \hline
Abundance & 8 Li mem with [Fe/H] values & \\
analysis & 2 Li mem with other elements estimates & \\ \hline \hline
\end{tabular}
\begin{tablenotes}
      \small
      \item * In the text we use ``Hot Candidate Members'' (HCMs) for the 
stars with (B$-$V$)_{0}$$<$0.35 for which we cannot use lithium as a membership criterion.
\end{tablenotes}

\label{uves-summary}
\end{center}
\end{threeparttable}
\end{table*}

As a first step, we searched for the presence of spectroscopic binaries in the sample of UVES stars. We identify six double-lined binaries (SB2) through their released CCFs: namely,
J08072516-4712522, J08073722-4705053, J08093589-4718525,
J08103996-4714428, J08105382-4719579, and J08115305-4654115. Those
systems were hence discarded from the sample analyzed for membership. 
Note that for one of the SB2 systems fundamental parameters are available.

\subsection{Radial velocity distribution \label{RV-mem}}
We have considered all UVES candidates with available RV and that
have not been identified as SB2 systems for the radial velocity analysis . This adds up to 62 stars.
Also, for the estimate of the RV membership, we have added 1.1~km/s to the RVs from UVES 
spectra to account for the systematic offset with respect to Giraffe
described in Sect.~\ref{verification-RV}.

Assuming that the UVES targets would be 
characterized by the same RV
distribution as the Giraffe targets, the analysis was performed
adopting the results of \citet{Jeffries14}; 
specifically, considering the two kinematic components identified in
that study, along with their peak
velocity and dispersion,
we defined as RV members all the stars with RVs in the interval between 14.9 and 22.3~km/s, corresponding to R$V_{2}$$\pm$2$\sigma_{2}$ of the broader distribution. Among the 62 UVES candidates, we have identified seven RV members and 55 stars whose RV values lie out from the R$V_{2}$$\pm$2$\sigma_{2}$ boundaries. Indeed, some of these stars can be binary systems that are
members of Gamma Velorum, however, hereafter, we will refer to this stars as RV nonmembers.

\subsection{Identification of the giant contaminants \label{logg}}
\begin{figure}
\hspace{-0.8cm}
\includegraphics[width=0.55\textwidth]{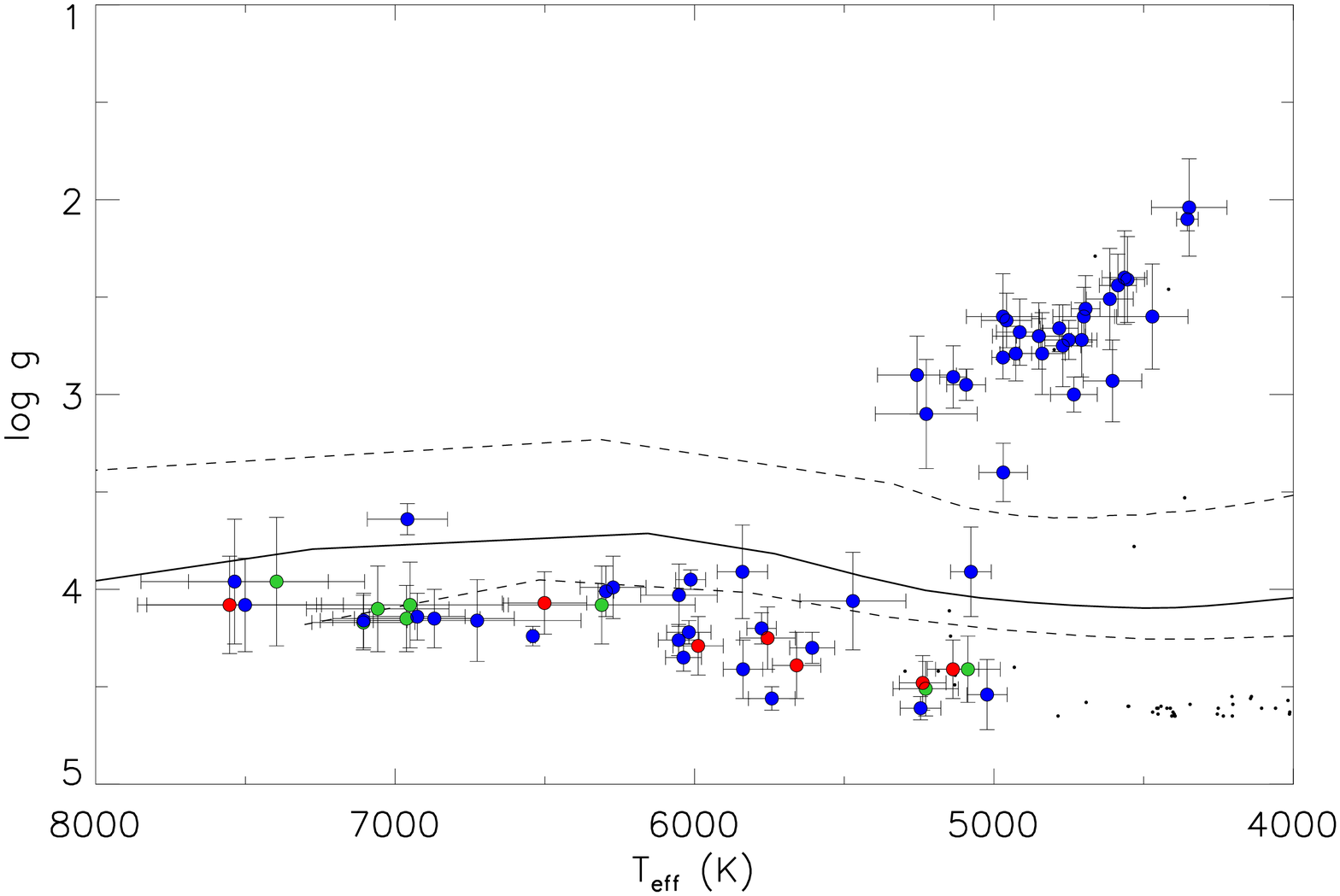}
\caption{Surface gravity versus effective temperature. The diagram allows us to identify the giant field stars in the UVES sample as the objects that lie above the 5~Myr isochrone (solid line) together with the 1 and 10 Myr (dashed lines) isochrones using \citet{Siess00} models. The different colors indicate RV members (red), RV nonmembers (blue), and stars excluded from the RV analysis (green). The black dots show the Giraffe members identified by \citet{Jeffries14}. Note that a few Giraffe Li members have gravity values below the 10 Myr isochrone. The diagram is limited to the temperature range relevant for the UVES targets.}
\label{logg_teff}
\end{figure}

The sequence of Gamma Velorum candidate members is easily identified in optical CMDs at 
magnitudes V$>$15 mag (see  \citealt{Jeffries09}), however, the UVES targets are restricted to 
the brighter part of the CMD where the sequence is heavily contaminated by field stars. 
In order to 
identify the population of evolved star contaminants, we plot in Fig.~\ref{logg_teff} 
the spectroscopic surface gravity as a function 
of $\rm T_{\rm eff}$ for the UVES targets along with the Giraffe targets identified as cluster members by \citet{Jeffries14}.
Stars are clearly divided in two groups: main-sequence and pre-main
sequence stars with $\log$~g between 4 and 5 dex, 
and giant stars with lower gravity values. In the figure, we also show the 5~Myr 
(solid line), 1 and 10 Myr (dashed lines) isochrones 
from \citet{Siess00} models for a 
metallicity of $Z_\odot$=0.01, close to the value of the solar metallicity, $Z$=0.012, adopted in the MARCS models. Based on this figure,
we conservatively consider all the UVES stars that lie above the 5~Myr isochrone as giant contaminants. 
Using this criterion we find that out of the 67 non-SB2 systems with $\log$~g 
determination, 37 
lie below the 5~Myr isochrone: 
the seven RV members, 22 candidates with RV not consistent
with that of the cluster, and eight stars without an RV estimate. 
The remaining 30 UVES stars, with $\log$~g typical of a giant
star will be discarded from further analysis. Not surprisingly, all these stars are RV nonmembers.

\subsection{Lithium members \label{Li-mem}}

\begin{figure*}
\hspace{-0.3cm}
\includegraphics[width=1.0\textwidth]{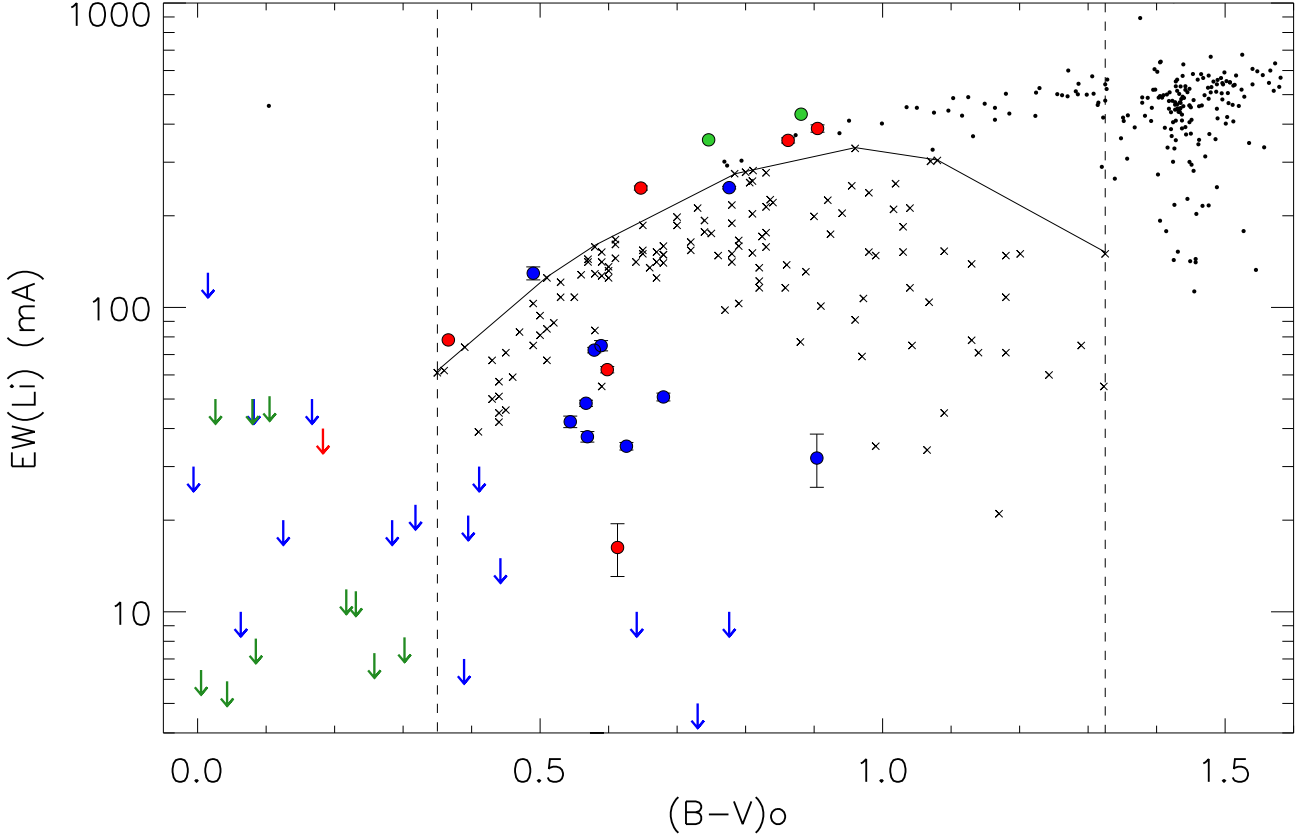}
\caption{Lithium EW as a function of the intrinsic color (B-V$)_{0}$. 
The red, blue, and green symbols (circles and arrows) represent the RV members, RV nonmembers, and stars with no RV estimate from the UVES sample, respectively. 
Most of the Li detections in UVES spectra have uncertainties associated with their EWs that are smaller than the data points.
The black dots show the Giraffe members identified by \citet{Jeffries14}. Their uncertainties, which have typical values of $\pm$10-20~m$\AA$, are not represented in the plot. Also, the uncertainties associated with the colors are not plotted, since their typical values of $\pm$0.02 are negligible. The solid line denotes the upper envelope of the Pleiades distribution (crosses; \citealt{Soderblom93,Jones96}). The dashed lines identify the color range 0.35~$<$~(B-V)$_0<$~1.33 spanned by the Pleiades members.}
\label{BV_Li}
\end{figure*}

As is well known, 
lithium is amongst the most useful membership indicator for young stars. 
In Fig.~$\ref{BV_Li}$, we show the EW(Li) 
as a function of the intrinsic color (B-V$)_0$ 
for the 44 UVES candidates that 
have not been rejected as SB2 systems or giant contaminants or
do not have any $\log$~g measurement. We derived the intrinsic B$-$V colors from the photometry reported in Jeffries et al. (2009) 
dereddened adopting the $E_{\rm{B-V}}$ estimated for Gamma Velorum by the same authors.
Along with the UVES stars, we also plot the 208
Giraffe targets classified as cluster members by \citet{Jeffries14}. 
Most of the Giraffe
targets are in the color range $0.7<$(B$-$V)$_0<1.3$ 
and have EW(Li)$>$200~m$\AA$.
Their distribution clearly defines the sequence of Li undepleted members. 
At (B-V)$_0>$~1.3, however, we observe a large dispersion in equivalent widths, 
indicating that a fraction of low-mass stars in the cluster have started
depleting lithium and hence suggesting a possible age dispersion (see
Jeffries et al. 2014; Franciosini et al. 2014, in prep.). 

In order to assess the membership of the UVES sources 
on the basis of the lithium content, we also use the available information 
for the members of the 
Pleiades cluster ($\sim$125-130~Myr; \citealt{Stauffer98}), similar to
the approach of \citet{James06}.
The comparison of the EWs(Li) of our sources with those of Pleiades members with
similar (B-V$)_0$  will allow us to identify the youngest targets, 
which are therefore the likely members of Gamma Velorum.
Among the UVES targets in the range of colors spanned by the Pleiades members
seven stars have EW(Li) higher than their Pleiades counterparts, 
since they lie above the upper envelope of the Pleiades Li-color distribution. 
One additional UVES target, with EW(Li)$>$200~m$\AA$, lies slightly below 
the upper envelope of the Pleiades. 
All of the other UVES stars have EW(Li)$<$100~m$\AA$ and are located
significantly below the Pleiades distribution. 
Most of these latter stars are RV nonmembers and their low lithium
suggests that they are not associated with the Gamma Velorum cluster and
are likely field contaminants. 
On the other hand, the seven stars with EW(Li) greater than the upper envelope 
of the Pleiades distribution are substantially younger than the Pleiades, 
thus are likely members of Gamma Velorum. The case of the star lying slightly below the upper envelope is less obvious and its membership needs to be further assessed on the basis of a CM diagram in Section~\ref{CMD-mem}. We note that
this star is an RV nonmember, hence a possible binary member.
To conclude,
we consider all of the stars with 0.35~$<$(B-V)$_0<$~1.33 lying above the upper envelope of the Pleiades distribution, plus the UVES target lying slightly below that limit as Li members/candidates. For stars with (B-V)$_0<$~0.35, we cannot use lithium as a membership criterion, but we flagged them as ``hot candidate members'' (hereafter, HCM). The membership of all these stars will be further checked in Section~\ref{CMD-mem}.

To summarize, the analysis of Li allows us to conclude that there are eight high-probability 
UVES Li members (four RV members, two RV nonmembers, and 
two stars without an RV estimate). Fifteen RV nonmembers appear to be older than the Pleiades counterparts, thus likely contaminants. Also two RV members have small EW(Li) and hence appear to be nonmembers based on their lithium content.
As for the hotter stars, there are 11 HCMs on the left side of the dashed line (one RV member, 
ten without RV estimate) that we will consider for further analysis and eight HCMs that are RV nonmembers, which will be rejected. 
Interestingly, we also note that one of the six SB2 systems (J08093589-4718525, $\#$46) has both components with EW(Li) larger than 100~m$\AA$, making it a possible member of the cluster.

\subsection{Color-magnitude diagram\label{CMD-mem}}
\begin{figure}
\hspace{-0.8cm}
\includegraphics[width=0.55\textwidth]{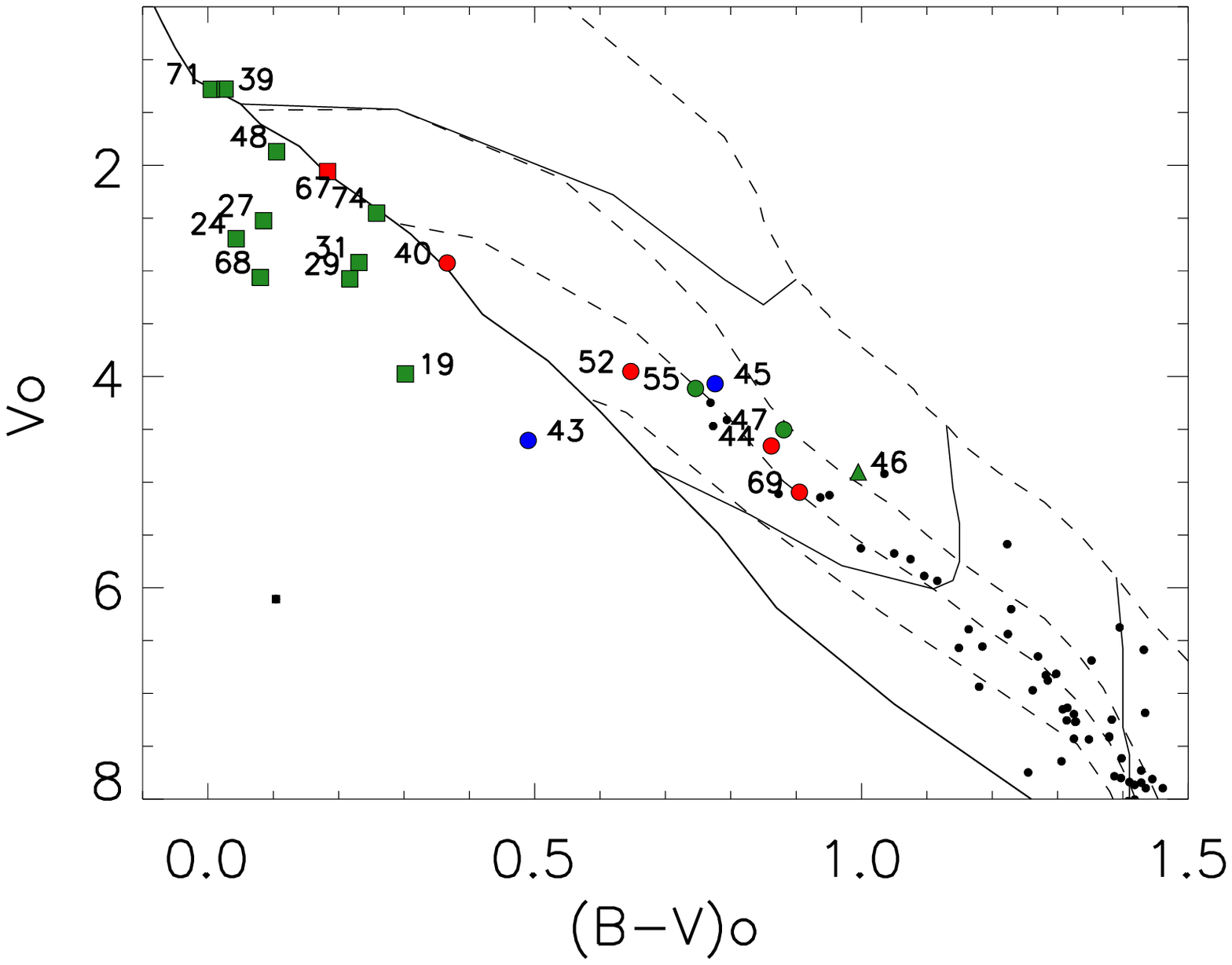} 
\caption{V$_0$ vs (B-V)$_0$ diagram of the UVES high-probability
members (circles), HCM (squares) and Li-rich binary system (triangle). Stars are color coded according to the RV membership. Each star is labelled according to the ID number given in Table~\ref{67parameters}. The Giraffe members from \citet{Jeffries14}} are shown as black dots. The solid and dashed lines are the evolutionary tracks for 0.5, 1 and 2 $M_{\sun}$ and isochrones for 1, 5, 10, 20~Myr and ZAMS from \citet{Siess00} for a chemical composition with $Z=$0.01. The diagram is limited to the color-magnitude range relevant for the UVES targets.
\label{CMD}
\end{figure}

The CMD is a helpful tool to confirm the reliability of our membership analysis and to provide 
some additional information about the HCMs for which we were not able to establish a secure 
membership based on their lithium content. Figure~\ref{CMD} shows the position of the eight UVES targets considered as Li members, plus the 11 HCMs and the Giraffe members from \citet{Jeffries14}; we made the diagram
using the photometry given by \citet{Jeffries09} and also released 
to the Gaia-ESO consortium. 
For stars not included in this compilation, we have used the photometry from the Tycho-2 catalog \citep{Hog00}. Both the distance modulus (DM=7.76) and reddening 
(E(B$-$V)=0.038) are taken from \citet{Jeffries09}. As expected, the majority of the stars fall 
in proximity or above the ZAMS in a sequence close to the 10~Myr isochrone, although a few 
outliers are present. Among the UVES
sample, there are six HCMs ($\#$19, 24, 27, 68, 31, 29) and one Li member ($\#$43) lying significantly 
below the ZAMS, i.e., more than the $\pm$0.3~mag spread in distance modulus found by \citet{Jeffries09} (uncertainties on photometry and extinction are negligible for these stars).
We will not consider these six HCMs for further analysis since they are likely field dwarfs. 
On the other hand, we note that star $\#$43 has an RV value slightly below ($\sim 1$ km/s) our
threshold for membership; however, its high EW(Li) and surface gravity 
($\log$~g=4.03~dex) are
consistent with those of other high-probability members. Star $\#$45, which lies slightly below the upper envelope of the Pleiades distribution in Fig.~\ref{BV_Li}, is consistent with the other UVES Li members and the Giraffe members in the CMD.
Thus, we include both these latter as likely members in the sample considered for the abundance analysis.

\subsection{Summary of the membership analysis \label{GVmem}}
In Table~\ref{UVES_targets}, we list the parameters of the 80 stars observed with UVES. 
In the last five columns, we give the membership status from the RV, surface gravity, 
EW(Li), and the position in the CM diagram: 
``M" stands for member, ``N" for field contaminant, and ``HCM" for hot candidate member.  
Table~\ref{uves-summary} summarizes the tally resulting from the analysis of the individual membership indicators.
In total, there are eight high-probability members, as indicated by their lithium, plus 
one Li-rich SB2 that can be considered a likely member. Among the high-probability members, four are also RV members. There are also five HCMs that satisfy the criteria for membership based on $\log$~g and the CMD. In total, we have 14 likely members. Note that the membership of two out of the seven 
RV members is not confirmed
by the lithium analysis, implying a contamination of about 30 \% in the 
RV sample.

\section{Abundance analysis of the members of Gamma Velorum \label{Discussion}}
\subsection{Iron abundance}
\begin{figure}
\hspace{-0.8cm}
\includegraphics[width=0.55\textwidth]{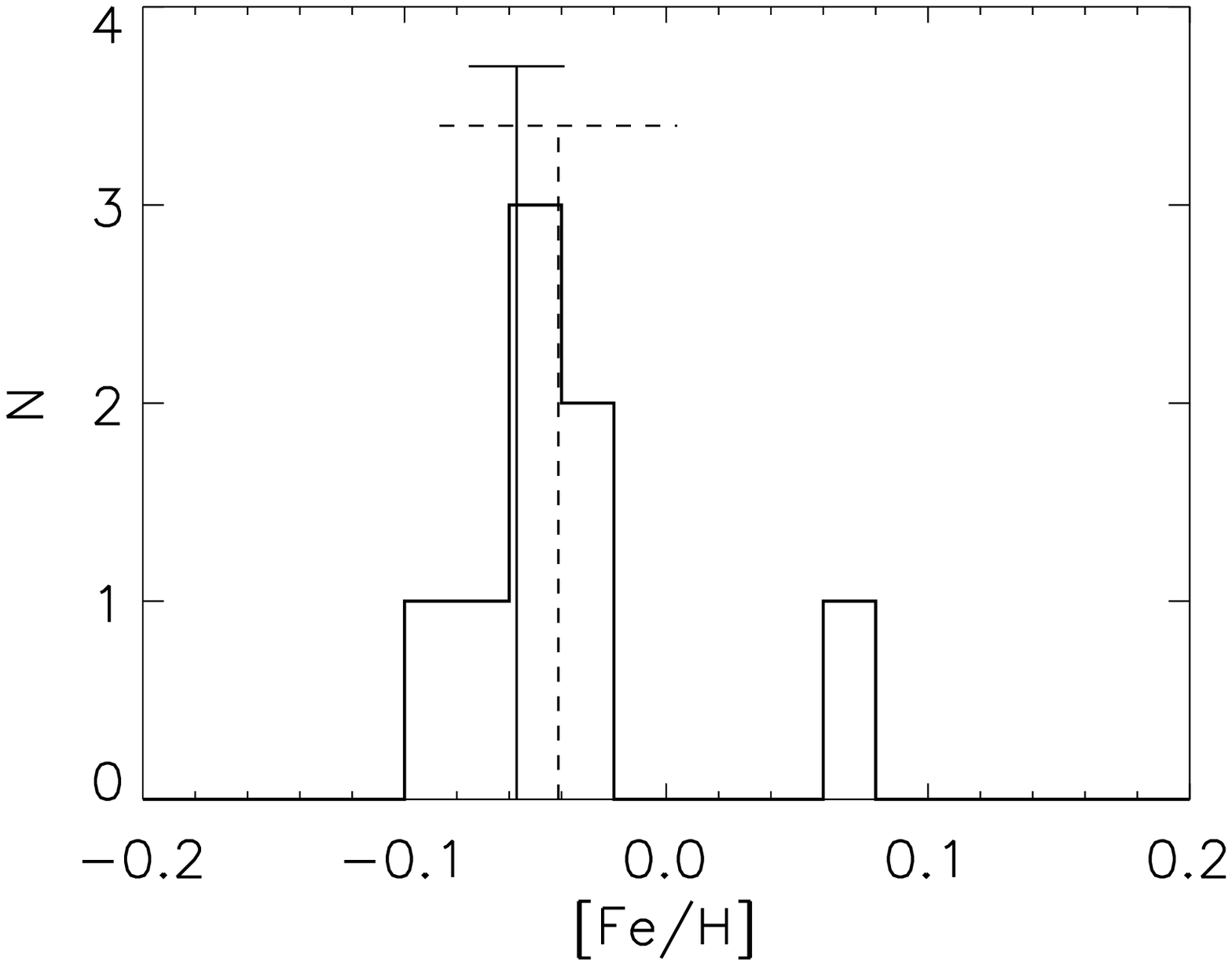}
\caption{Distribution of the iron abundance of the eight UVES high-probability cluster members. The mean values are $<$[Fe/H]$>$=$-$0.04$\pm$0.05~dex (dashed line) and $<$[Fe/H]$>$=$-$0.057$\pm$0.018~dex (solid line) discarding the star \#52 with [Fe/H]$=0.07$~dex.}
\label{iron}
\end{figure}

Based on the eight UVES high-probability members, 
we obtain the [Fe/H] distribution of Gamma Velorum shown
in Fig.~$\ref{iron}$. The mean iron abundance is $<$[Fe/H]$>=-$0.04$\pm$0.05~dex, 
where the error corresponds to 1$\sigma$ of the distribution. 
We recall that this [Fe/H] value refers to a solar value of 
$\log$~n(Fe)$=$7.45 \citep{Grevesse07}. 
While seven of the eight stars have abundances in the narrow range $-$0.1 to $-$0.03 dex, the difference between the metallicity of J08095427-4721419 ($\#$52; [Fe/H]=$+$0.07~dex) and the mean is larger than $\sim$2$\sigma$. 

The membership of this star 
is based on the gravity, the presence of photospheric Li,
and on an RV consistent with that of the cluster. Other indirect supports come from the relatively high rotational velocity typical of young stars, a proper motion consistent
with the other members of the cluster and high level of X-ray emission
\citep{Jeffries09}. Furthermore, the star exhibits an IR excess at 24~$\mu$m that suggests the presence of a debris disk \citep{Hernandez08}. The possible origin of the high-iron abundance of J08095427-4721419 is discussed in Sect.~$\ref{metal-rich}$. If we excluded this member with a peculiar high metallicity, we 
would obtain a mean iron abundance of $-$0.057$\pm$0.018~dex.
These results indicate that the members of Gamma Velorum have a slightly subsolar iron abundance with a small dispersion. 
The mean iron abundance is compatible with that derived in other young 
open clusters of the solar neighborhood \citep{Biazzo11a}, while the small 
scatter suggests a homogeneous iron abundance in Gamma Velorum. 
Considering the two kinematic groups, we note that among the eight members only one star (J08110285$-$4724405, $\#$69) is more likely 
associated with Population A in
Jeffries et al. (2014), while the remaining seven ones more likely
belong to Population B. Hence, based on the UVES targets,
we cannot make a comparative analysis of the abundances in terms of 
the two RV populations. 

\begin{figure}
\hspace{-0.8cm}
\includegraphics[width=0.55\textwidth]{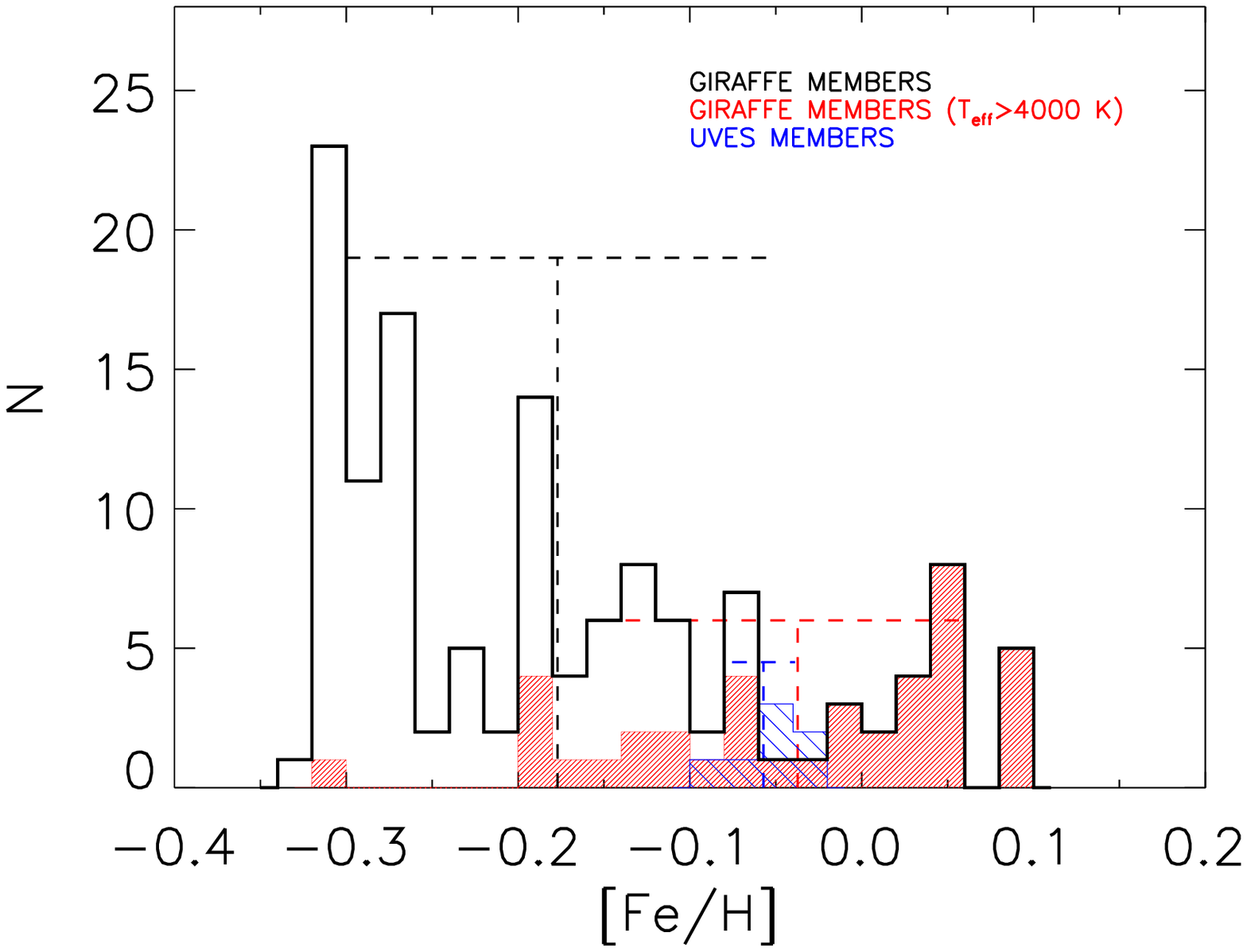}
\caption{Iron abundance of the Giraffe targets identified by \citet{Jeffries14} as members (solid histogram). The other histograms show the iron abundance og the Giraffe members with $\rm T_{\rm eff}$$>$4000~K (red) and that of the UVES members (blue).}
\label{iron_distr_Giraffe}
\end{figure}

The much larger number of Giraffe members and the richer statistics allow for
a more general study of the iron abundance distribution in Gamma Velorum. 
In Fig.~\ref{iron_distr_Giraffe}, we show the iron abundance of the Giraffe 
members from \citet{Jeffries14} with v$\sin i$~$\leq$~50~km/s. 
The latter constraint is necessary 
because in the spectra of fast rotating stars
the blending of the absorption features makes the iron abundance 
estimate highly uncertain. We find that $<$[Fe/H]$>$$_{\rm Giraffe}$~=~$-$0.18~$\pm$~0.12~dex is consistent within the errors, though somewhat lower, with the mean iron abundance based on the
analysis of UVES spectra (Fig.~\ref{iron}). 
The different mean
iron abundance and the broader width of the distribution 
(from $-$0.34 to $+$0.13~dex) are partly due to the lower resolution of the Giraffe spectra and also to the intrinsic difficulty of the analysis of cool stars. 
If we restrict the Giraffe sample to stars with $\rm T_{\rm eff}$~$>$~4000~K ($\sim$25$\%$ of the objects), we derive a mean value of $<$[Fe/H]$>$$_{\rm Giraffe}$~=~$-$0.04~$\pm$~0.10~dex, with a greater similarity to that of the UVES sample. Note that
a few stars with high metallicity are present in the Giraffe distribution, comparable to that of star J08095427$-$4721419. Whereas the
larger typical uncertainties of the Giraffe determinations certainly
contribute to broadening the [Fe/H] distribution, we cannot exclude the possibility that 
the Giraffe sample also contains a number of as yet unidentified metal-rich 
outliers, as well as field contaminants or other difficult stars, for example binaries, which could affect the distribution.

\begin{figure}
\hspace{-0.8cm}
\includegraphics[width=0.55\textwidth]{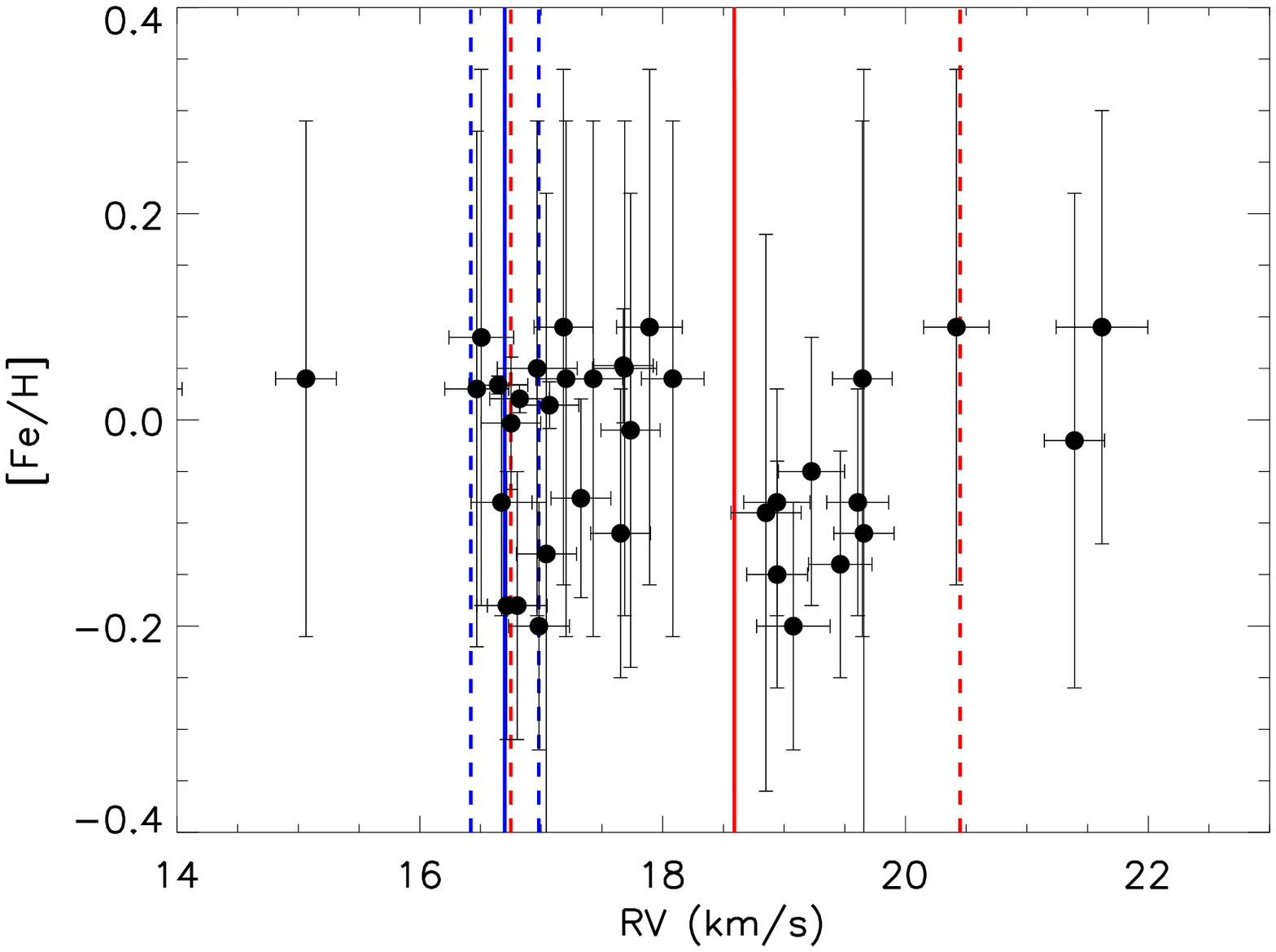}
\caption{Iron abundance of the Giraffe targets identified by \citet{Jeffries14} as members} and with $\rm T_{\rm eff}$$>$4000~K as a function of their RVs. The red and blue solid lines mark the central RV of each young population identified by Jeffries et al. (2014). 
The dashed lines corresponds to $\pm$1$\sigma$ central value.
\label{Giraffe_pop_metal}
\end{figure}

Finally, considering the Giraffe members with $\rm T_{\rm eff}$~$>$~4000~K, we plot in Fig.~\ref{Giraffe_pop_metal} their iron abundances as a function of the RVs. Although a group of stars with RV values ranging between 18.5 and 20 km/s,
likely belonging to Population B, have lower iron abundances with
respect to the other stars, there is a significant scatter among the 
stars of the two populations without hints of abundance separations.
Thus, we conclude that the two groups likely have the same iron abundance.


\subsection{Other elements}
The Gaia-ESO Survey has released for Gamma Velorum
the abundance of elements other than 
iron for stars with v$\sin i$$<$20 km/s. Unfortunately, this limits the 
analysis to two confirmed members only: 
J08095427$-$4721419 (\#52) and J08093304$-$4737066 (\#45).
The elemental abundances are listed in Table~$\ref{elemtab}$.

We see that for the cooler star (\#45) the abundances are within $\pm$0.1~dex of the solar values, with the only exception being calcium and nickel, which
are enhanced and subsolar, respectively. 
The warmer star (\#52) shows abundances significantly larger than solar and
very unusual for the solar neighborhood for most of the analyzed elements. 

These abundances may, in principle, shed light on the reasons for the high metallicity found in star \#52 given that it is a metal-rich star; however, the errors are much larger than those obtained for $\#$45, with the exception of iron. Hence, we cannot attempt any 
conclusions based on the abundance ratios.

\begin{table}
\begin{center}
\caption{Elemental abundances}
\footnotesize
\begin{tabular}{rcc}
\hline\hline
& J08093304$-$4737066 & J08095427$-$4721419
 \\
& $\#$45 & $\#$52 \\
\hline\hline
   $\rm T_{\rm eff}^{\rm spec}$ (K)& 5471 & 5756   \\
   $[$Fe/H$]$	& $-$0.06$\pm$0.12 & $+$0.07$\pm$0.07  \\
   $[$Si/Fe$]$	& $-$0.10$\pm$0.08 & $+$0.18$\pm$0.20  \\
   $[$Mg/Fe$]$	& $+$0.02$\pm$0.14 & $+$0.30$\pm$0.40  \\
   $[$$<$Ti$>$/Fe$]$ & $-$0.04$\pm$0.07 & $+$0.46$\pm$0.29 \\
   $[$Ca/Fe$]$	& $+$0.18$\pm$0.03 & $+$0.08$\pm$0.40 \\
   $[$Cr/Fe$]$ & $+$0.06$\pm$0.03 & $+$0.47$\pm$0.22 \\
   $[$Ni/Fe$]$	& $-$0.18$\pm$0.05 & $+$0.08$\pm$0.52 \\
\hline \hline  
\end{tabular}
\label{elemtab}\\
\end{center}
\end{table}

\subsection{The metal rich star J08095427$-$4721419 \label{metal-rich}}
J08095427$-$4721419, one of the eight high-probability members of Gamma Velorum,
 has an iron abundance of [Fe/H]~=~$+$0.07~$\pm$~0.07~dex. 
We also notice that different analysis methods within the Gaia-ESO
consortium have derived similar stellar parameters and enhanced metallicity (see Table~\ref{U51512}). The position in the CMD reinforces the quality of the atmospheric parameters. Thus, we assume that this star is genuinely more metal-rich than other cluster members.

\begin{table}
\begin{center}
\begin{threeparttable}
\caption{Fundamental parameters of J08095427$-$4721419.}
\footnotesize
\begin{tabular}{cccc}
\hline\hline
& T$_{eff}$ & $\log$~g & $[$Fe/H$]$  \\
& (K) & (dex) & (dex) \\
\hline\hline
WG11 {\tiny average} & 5756 $\pm$ 93 & 4.25 $\pm$ 0.16 & $+$0.07 $\pm$ 0.07  \\
WG12 {\tiny average} & 5864 $\pm$ 112 & 4.41 $\pm$ 0.08 & $+$0.14 $\pm$ 0.09 \\
WG1$2^{a}$ & 5944 $\pm$ 57 & 4.45 $\pm$ 0.10 & $+$0.20 $\pm$ 0.13 \\
WG1$2^{b}$ & 5785 $\pm$ 56 & 4.37 $\pm$ 0.13 & $+$0.08 $\pm$ 0.11 \\
\hline \hline  
\end{tabular}
\begin{tablenotes}
      \small
      \item a) Iron EWs method.
      \item b) Comparison with a library of standard star spectra.
\end{tablenotes}
\label{U51512}
\end{threeparttable}
\end{center}
\end{table}

Based on statistical considerations, one would expect to detect one 2$\sigma$ outlier in a sample containing
more than 20 stars; hence, the probability of having one outlier
out of eight members is rather small, although not negligible.
Under the assumption that the star is a genuine metal-rich cluster member, we 
propose the following scenario to explain it.

First, we believe that the chemical enrichment due to the explosion of a nearby SN is unlikely since it should have enriched the whole cloud and other members. A more likely process is the accretion of circumstellar rocky material onto the star that is mixed in the stellar convective envelope causing an overall metallicity enhancement \citep{Laughlin97}. If the star is cool and young, its extended outer convection zone  will effectively mix the accreted material with only a minimal metallicity enhancement. On the other hand, if the star is mostly radiative with a thin convective layer, the pollution could be much more important, leading to observable consequences. 

Applying the scenario proposed by \citet{Laughlin97}, a solar-type star, like J08095427-4721419, starts its PMS contraction with a fully convective structure, but after $\sim$2~Myr a radiative core appears that grows in mass as the star ages, shrinking the outer convective layers. Such a star maintains a thick convective envelope until about 10 Myr. For later-type stars, the growth of the radiative core takes more time and the final thickness of the convective layer is larger; in earlier-type stars, the radiative core develops quickly until a fully radiative configuration is reached. Circumstellar disks are found in most of young stellar objects and generally they accrete onto the central star during the first 10 Myr when their internal structure is still mainly convective. The condensation of heavy elements could lead to the formation of rocky blocks or planets, however, which prevents a quick accretion of this material during the time when the star is mainly convective. In the last decade, several surveys have shown that a great number of extrasolar planets have surprisingly small orbits, suggesting that after their formation significant orbital migration takes place in the protoplanetary system. A possible outcome of this inward migration is that part of the planetary material reaches the central star even after the completion of the main accretion phases.

\subsection{A quantitative estimate of the effects of rocky material accretion on the mainly radiative PMS star J08095427$-$4721419}
Using $\rm T_{\rm eff}=$5756~K and $L_{\rm bol}=$2.5~L$_\odot$ for J08095427$-$4721419 and the Siess models for a subsolar metallicity \citep{Siess00}, we derive a stellar mass of $\sim$1.3~$M_{\sun}$ and an age of $\sim$15-16~Myr. 
Thus, this star appears somewhat 
older than the average age of the cluster, but consistent with the age dispersion found by Jeffries et al. (2009, 2014).
As we mentioned above, 
the {\it Spitzer} data show evidence for the presence of a debris disk \citep{Hernandez08}. Thus, we  can imagine that part of the circumstellar matter has condensed into hydrogen-depleted rocks, or even planets, and that this rocky material has recently accreted onto the star. The Siess models also predict that such a star is almost fully radiative, but that about 5~Myr in the past it had a thin convective layer of $\sim$0.05~$M_{\sun}$. We now estimate the mass of heavy elements (expressed in Earth masses, $M_{\oplus}$$\sim$3$\times$1$0^{-6}$~$M_{\sun}$) that must have been accreted onto the star during the last 5~Myr to produce an iron enhancement similar to that observed in J08095427$-$4721419.
For this purpose, we assume that the accreted material is mixed in a convective region containing 0.05~$M_{\sun}$. We also assume that the star has an initial iron abundance equal to the average value of the other members
([Fe/H$]_{\rm init}$$=$$-$0.057~dex). Now, the rocky material, being hydrogen-depleted, has a mass ratio of metals $Z_{p}$=1, and we assume that such a rocky mass has the heavy element distribution as the solar mix given by \citet{Grevesse07}.
Hence, a mixing of 50, 60, and 70~$M_{\oplus}$ in the 0.05~$M_{\sun}$ convective layer would be enough to produce a metallicity variation ($\Delta$Z) of 3.0$\times$1$0^{-3}$, 3.6$\times$1$0^{-3}$, and 4.1$\times$1$0^{-3}$, corresponding to a final iron abundance of  [Fe/H$]_{fin}=+$0.05, $+$0.07 and $+$0.08 dex. This is just the right amount needed to explain the observed iron abundance. The same effect could be achieved by the accretion of two Jupiters ($M_{J}$$=$0.001~$M_{\sun}$) with a metallicity $Z=$0.1.

We recall that the mass of heavy elements currently contained in the planets of the solar system is estimated to be in the range 60-120$M_{\oplus}$ \citep{Wuchterl00}. This number is also in the range of heavy-element mass for exoplanets found by \citet{Miller11}. Thus, our estimate of the accreted mass is consistent with these numbers. On the other hand, we have found only one star with a significant metallicity enhancement, whereas the presence of circumstellar disks is a frequent phenomenon around young stars and more than just one example should have been found in Gamma Velorum. Our proposed scenario requires the refinement of several factors. First, the star must be of the right mass to possess a convective region that shrinks significantly while contracting. Then, this star should have the right amount of mass in the convective layer since otherwise the accreted metals could be too diluted or, conversely, enhanced with respect to the observed abundance. Third, the accretion episode must have occurred only after the star has had time to contract significantly for the retreat of the convection layer and this requires several Myr. The fact that J08095427-4721419 is a bit older than the other members of the cluster, judging from its isochronal age, supports our interpretation. A similar scenario has recently been suggested by \citet{Theado12} in the context of the predicted modifications of the light element abundances of accreting exoplanet-host stars. Although the case of J08095427$-$4721419 is the only one found so far in Gamma Velorum, we should also mention other examples of metal-rich stars in other solar or subsolar young clusters and star forming regions (e.g., \citealt{Wilden02,Biazzo11a}).

\section{Conclusions \label{conclusions}}
In this paper, we have made use of the dataset provided by the Gaia-ESO Survey to identify the Gamma Velorum members in the UVES sample
and to study their elemental abundances, in particular
to derive the mean cluster metallicity.
The main findings can be summarized as follows:\\

\noindent
i) The main result of this paper is the first metallicity estimation of the Gamma Velorum cluster. We find that
it has a slightly subsolar mean iron abundance: $<$[Fe/H]$>$$=$$-$0.057~$\pm$~0.018~dex, if we exclude the metal-rich star J08095427$-$4721419. The analysis of other heavy element ($\alpha$, iron peak, etc.) abundances for two members is not conclusive, given the very large uncertainties.
This is the first estimate of the metallicity of Gamma Velorum. When compared
with the metallicity of other clusters belonging to the Vela complex and
observed by the Gaia-ESO Survey (but whose analysis has not yet been
completed), it will possibly allow us to put constraints on the star formation
process in the complex.\\

\noindent
ii) In order to determine the iron content of the cluster, we performed the membership analysis on the whole sample of targets. Using RVs, surface gravity and the presence of Li in the stellar atmospheres, we have identified eight high-probability members.
We have also detected one SB2 system (J08093589-4718525, $\#$46) whose components both display a strong lithium line. This spectroscopic binary could be considered a likely member. Furthermore, we have identified five hot-candidate members of the cluster based on their position in the CMD.\\

\noindent
iii) We have found a metal-rich member, J08095427$-$4721419. Its mass ($\sim$1.3~M$_\odot$) and age ($\sim$15~Myr) are consistent with an internal structure characterized by a thin convective envelope. We have suggested a scenario to account for the observed increase of the atmospheric abundances based on the accretion of $\sim$60~$M_{\oplus}$ of rocky hydrogen-depleted material onto the star. \\

\noindent
iv) The average metallicity derived from the Giraffe sample
is similar to the average for UVES. A few metal-rich stars are also present
in the Giraffe sample, but their presence may be due
to the larger uncertainties and dispersion. Based on Giraffe sample,
no major difference in the [Fe/H] distribution is found for
the two kinematic population identified by Jeffries et al. (2014).\\

On the more technical aspects, the comparison of the parameters from the UVES
and Giraffe analysis of the same stars observed in Gamma Velorum can be summarized as follows:\\

\noindent
i) The Giraffe radial velocities of the first release are systematically higher with respect to the UVES values by 1.1$\pm$0.4~km/s (see \citealt{Sacco14} for a detailed discussion).\\

\noindent
ii) The stellar parameters $\log$~g and T$_{\rm eff}$ and the measured
Li EWs are generally in good agreement, although small 
discrepancies are present for $T_{eff}$$>$5500~K and  EW(Li)$<$30~m$\AA$.\\

\noindent
iii) There is a reasonable agreement between the iron abundances derived using Giraffe and UVES, but the intrinsic dispersion of the latter is significantly smaller.\\

\begin{acknowledgement}
We acknowledge the support from INAF and Ministero dell'Istruzione, dell'Universit\`{a} e della Ricerca (MIUR) in the form of the grant ``Premiale VLT 2012''. We also acknowledge the financial support from ``Programme National de Cosmologie and Galaxies'' (PNCG) of CNRS/INSU, France. The results presented here benefited from discussions in three Gaia-ESO workshops supported by the ESF (European Science Foundation) through the GREAT (Gaia Research for European Astronomy Training) Research Network Program (Science meetings 3855, 4127 and 4415). T.B. was funded by grant No. 621-2009-3911 from The Swedish Research Council. This research has made use of the SIMBAD database, operated at CDS, Strasbourg, France. E.J.A. acknowledges financial support from the ``Ministerio de Econom\'{i}a y Competitividad'' of Spain through grant AYA2010-17631. S.G.S, EDM, and V.Zh.A. acknowledge support from the Funda\c{c}\~{a}o para a Ci\^{e}ncia e Tecnologia (Portugal) in the form of grants SFRH/BPD/47611/2008, SFRH/BPD/76606/2011, SFRH/BPD/70574/2010, respectively.
\end{acknowledgement}

\bibliographystyle{aa} 
\bibliography{bibliography}

\end{document}